\documentstyle[preprint,aps]{revtex}
\begin{document}
\tightenlines
\title{
The description for the  spin
polarizabilities of hadrons \\ based on the covariant Lagrangian }
\author{S.A. Belousova, N.V.  Maksimenko}
\address{ Gomel State
University, Sovetskay street, 102, Gomel, Belarus, 246699;\\ E-mail:
belousova@gsu.unibel.by }
\date{\today}
\maketitle
\begin{abstract}
On the basis of the correspondence principle between the
relativistic moving medium electrodynamics and relativistic quantum
field theory the covariant Lagrangian of the electromagnetic field
interaction with the polarized spin particles have been obtained.
This Lagrangian satisfies the main relativistic quantum field theory
requirements and contains four independent covariant spin
structures, which have particular physical meaning. It is
shown that the spin polarizabilities  give the contribution
to the amplitude for Compton scattering on the spin-1/2 hadron in
the  ${\cal O}(\omega^3)$.
\end{abstract}
\narrowtext
In the expansion of the amplitude for Compton
 scattering in powers of the incident photon energy $\omega $, the
terms of ${\cal O}(1)$ and ${\cal O}(\omega)$, depend only
on the mass of the nucleon $m$, its electric charge $¥$ and the
anomalous magnetic moment $\mu$  (see, for example, Refs.
\cite{1,2}). By virtue of the low-energy theorem in the ${\cal
O}(\omega ^2)$ the amplitude for spin-0 and -1/2 hadrons depends on
their internal degrees of freedom, which are determined by the
fundamental parameters such as the electric ($\alpha$) and
magnetic ($\beta$) polarizabilities \cite{1,2}.  In turn, at the
${\cal O}(\omega ^3)$ the effective Lagrangian discribing the
photon-nucleon interaction and, as the result the amplitude for
Compton scattering are determined by gyrations, or so-called spin
polarizabilities, the new electromagnetic characteristics
\cite{3,3a,4,4a,5,6,7}. These characteristics are immediately
connected to the spin properties of  hadrons as  a composite
particles.

The classic process for the investigation of these features of
the photon-hadron interaction is the Compton scattering of photons
whose energies are less then resonance region. Nevertheless data
about spin polarizabilities can be extracted from other
electrodynamic processes (see, for example Ref. \cite{8}).

The determination of the hadron  polarizability contributions to
the amplitudes of QED processes is sequentially  carried
out by the effective Lagrangians of the electromagnetic field
interaction with the hadron as a composite particle. In the
nonrelativistic
electrodynamics the such kind of Lagrangian is rather
well determined.
On the other side in the relativistic QED when hadrons look like
 bound states, due to the kinematic relativistic effects,
 the interpretation of polarizabilities is
ambiguous  \cite{2,7}.

In the present report the correspondence principle
between relativistic moving medium electrodynamics and
relativistic quantum field theory will be sequentially used
for  the covariant Lagrangian construction
of  the electromagnetic field interaction with polarized spin
hadrons.

First of all we define the Lagrangian
of  the photon-hadron interaction taking
into account the electric and magnetic polarizabilities.

In the nonrelativistic case, the Hamiltonian of the interaction
  of the isotropicly-gyrotropic medium
  looks like \cite{9,10}:
\begin{equation}
\label{h1}
H_I=-2\pi \left (\bf
 P\bf E+\bf M\bf H \right),
\end{equation}
In this case vectors of
polarization $\bf{P}$ and magnetization $\bf{M}$ look like
\cite{9}
\begin{equation}
\label{h22}
\bf{P}=\hat \alpha \bf{E},
\end{equation}
\begin{equation}
\label{h23}
\bf{M}=\hat \beta \bf
{H},
\end{equation}
where $\hat \alpha$ and $\hat \beta$ are the tensors of electric and
magnetic polarizabilities, $\bf {E}$ and $\bf {H}$
are  the strengths vectors of electric and magnetic fields.
If the polarizability tensors in Eqs.(\ref{h22}) and (\ref{h23})
depend on the spin operator components  $\hat S_i$, then due to
commutation relations of these operators the Compton scattering
amplitude for spin-0, -1/2 and -1  hadrons is defined by the
following expression:
\begin{eqnarray}
\label{d12}
\frac 1{8\pi m}T_{fi}^{NB, nospin}&=&
\omega ^2\chi^{+}\{\alpha
_0~\bf{e}^{~\prime*} \cdot \bf{e}+i\alpha _1~\hat{
\bf{S}} \cdot \bf{e} \times \bf{e}^{~\prime*} +
\nonumber \\
&&+\alpha _2\{\hat {\bf{S}}\cdot \bf{e}^{~\prime*},~\hat
{\bf{S}}\cdot \bf{e}\}+
\beta_0~\bf{s} \cdot \bf{s}^{~\prime*}+
\nonumber \\
&&+i\beta_1~\hat
{\bf{S}}\cdot \bf{s}^{~\prime*}\times \bf{s}+
\beta_2\{\hat {\bf{S}}\cdot \bf{s},~\hat{ \bf{S}}\cdot
\bf{s}^{~\prime*}\}
\}\chi.
\end{eqnarray}
Here $\chi$ and $\chi^+$ are the  spin wave functions of the
particles, $\omega$ is the photon energy, $\bf{n}$ ($\bf{e}$) and
$\bf{n}^{~\prime}$ ($\bf{e}^{~\prime}$) are unit (polarization) vectors of
incident and scattered photons respectively,
 $ \bf {s}=\bf {n}\times  \bf {e};~\bf {s}^{\prime* }= \bf
{n}^{\prime }\times \bf {e}^{\prime* }.$
We assume $c=\hbar=1$.

Due to the crossing symmetry requirement the amplitude
(\ref{d12}) for the spin-0 and -1/2 hadrons are determined by
the parameters $\alpha_0$ and $\beta_0$.
The scattering amplitude for vector hadrons together with
the ordinary polarizabilities $\alpha_0$ and $\beta_0$ contains
spin ones $\alpha_2$ and  $\beta_2$, while the quantities
$\alpha_1$ and
$\beta_1$ are equal to zero because of the crossing symmetry.

Let us define now the effective Lagrangian
in the relativistic QED for spin-0, -1/2 and -1 hadrons
taking into consideration their usual electric  and magnetic
polarizabilities.
According to the relativistic moving medium electrodynamics
the effective Lagrangian  is \cite{10}:
\begin{equation}
\label{h15}
L_{eff}^{pol}=2\pi
\left\{e^\mu \alpha_{\mu\nu} e^\nu + h^\mu \beta_{\mu\nu} h^\nu
\right\}.
\end{equation}
In this expression $e_\mu=
F_{\mu\nu}U^\nu,~~h_\mu=\tilde F_{\mu\nu}U^\nu,~~ \tilde
F_{\mu\nu}=\frac{1}{2} \varepsilon _{\mu \nu \rho \sigma
}F^{\rho\sigma }$, where $F^{\mu\nu}$ and $\tilde F^{\mu\nu}$
are the electromagnetic field tensor and dual one respectively
($ F^{\it 0i}=-E^{\it i}, F^{\it
ik}=-\varepsilon_{\it ikl}\ H^{\it l} $), $\alpha_{\mu\nu}$ and
$\beta_{\mu\nu}$ are tensors, defining the polarizabilities of a
medium at rest, $U_\mu$ is the 4-velocity of the medium and   the
Levi-Chevita antisymmetric tensor $\varepsilon _{\mu \nu\rho \sigma
}$ is fixed by the condition $\varepsilon^{0123}=1$.

Following to the non-relativistic quantum field theory we
assumed that tensors $\alpha_{\mu\nu}$ and $\beta_{\mu\nu}$
depend on the momentum operator and Pauli-Lubansky's vector
operator:
\begin{eqnarray}
\alpha_{\mu\nu}=\alpha_{\mu\nu}(\hat W_\mu, \hat
p_\mu),
\nonumber \\
\beta_{\mu\nu}=\beta_{\mu\nu}(\hat W_\mu, \hat p_\mu),
\nonumber \\
\hat{ W}_{\mu}= \frac 1{2m} \varepsilon_{\mu \nu \rho
\sigma}~J^{[\nu \rho]}~ \hat {p}^{\sigma}.
\end{eqnarray}
Here $J^{[\nu\rho]}$ are
generators of the  Lorentz group.

Using the commutation relations for  the operators $\hat
W_\mu$ and $\hat p_\mu$ in the framework of the quantum-mechanical
Poincare group \cite{11} the expansions for spin-1/2 particles read:
  \begin{equation}
\label{h16}
\alpha_{\mu \nu }= {\alpha _0}g_{\mu \nu }+ {\alpha_1}\varepsilon
_{\mu \nu \rho \sigma } \hat{W}^\rho \hat {p}^\sigma ,
\end{equation}
\begin{equation}
\label{h17} \beta _{\mu \nu }=
{\beta _0}g_{\mu \nu }+ {\beta_1}\varepsilon _{\mu \nu \rho \sigma
}\hat{W}^\rho \hat {p}^\sigma .
\end{equation}

From Eqs. (\ref{h16}) and (\ref{h17}) we can see that when $\alpha _1$
and $\beta _1$ are equal to zero, the amplitude
satisfies the crossing symmetry ({\it i.e.}
at the ${\cal O}(\omega ^2)$
the Lagrangian (\ref{h15}) is defined by $\alpha _0$ and $\beta _0$
).

Going from the Lagrangian  ~(\ref{h15}) to the
field theory Lagrangian by means of the correspondence principle,
we obtain \cite {6,12}:
\begin{enumerate}
\item
for spin-0 particles
\begin{eqnarray}
\label{h20}
\displaystyle
L^{pol}_{eff}&=&
\displaystyle
\frac {\pi}{m}[ ( \partial _\mu \varphi
^{+})(\partial _\nu \varphi) +(\partial _\mu \varphi)(\partial _\nu
\varphi ^{+})-
\nonumber \\[0.3cm]
&&
\displaystyle
- ( \varphi
^{+}\partial _\mu \partial _\nu \varphi +\varphi \partial _\mu
 \partial _\nu \varphi ^{+})] K^{\mu \nu },
\end{eqnarray}
\item
for spin-1/2 particles
 \begin{equation}
\label{h21}
L^{pol}_{eff}=-\frac {i\pi}{m} \left(\stackrel{-}{\psi }
{\gamma^\mu}\stackrel{\leftrightarrow }{\partial _\nu} \psi \right)
K_{\mu}^{\nu },
\end{equation}
\end{enumerate}
where $ K_{\mu}^{\nu }= \alpha_0 F_{\mu
 \rho} F^{\rho \nu} + \beta_0 \tilde {F}_{\mu\rho} \tilde {F}^{\rho
\nu},$
$~~\stackrel{\leftrightarrow }{\partial
_\nu }=\stackrel{\leftarrow }{\partial
_\nu }-\stackrel{\rightarrow } {\partial
_\nu },$ $~~\varphi$ and $\psi$ are the wave functions of
spin-0 and spin-1/2 particles.
We have obtained  the amplitude for Compton scattering in the ${\cal
O}(\omega^2)$, taking into account the Lagrangian (\ref{h21}):
\begin{eqnarray}
\label{h22a}
\frac 1{8\pi m}T_{fi}^{NB,
nospin}=\chi ^{+}\omega _1\omega _2 [\alpha _{E}~ \bf {e}^{\prime *
}\cdot \bf {e}+ \beta _M~ \bf {s}^{\prime * }\cdot \bf{s}]\chi,
\end{eqnarray} where $ \bf {s}=\bf {n}\times  \bf {e};~\bf
{s}^{\prime* }= \bf {n}^{\prime }\times \bf {e}^{\prime* } ,  $
$\omega _1$ and $\omega _2$ are the energies of the incident and
scattered photons.

In order to define the effective Lagrangian with account of
  the nucleon spin polarizabilities, is used Eq. (\ref{h1}),
where polarization and magnetization vectors ($\bf P$ and $\bf M$)
are determined as follows
\begin{equation}
\label{22}
\bf{P}=\hat
\alpha \bf{E} +{\hat {\gamma}}^{\prime}_E \left[ {\bf {\nabla}}
 \bf {E}\right] ,
\end{equation}
\begin{equation}
\label{23}
\bf{M}=\hat \beta \bf {H} +{\hat {\gamma}}^{\prime}_M \left[ {\bf
{\nabla}} \bf {H}\right] .
\end{equation}

Then the following
Lagrangian of the electromagnetic field interaction with  the
moving medium  is obtained:
\begin{eqnarray}
\label{h24}
{L_I}^{eff}&=&2\pi
\{[e_\mu \alpha^{\mu\nu} e_\nu + h_\mu \beta^{\mu\nu} h_\nu]
-
\nonumber \\
&&
-[{({\gamma}^{\prime}_E)}_{\mu\nu} e^{\mu} (U\partial)
h^{\nu}+
{({\gamma}^{\prime}_M)}_{\mu \nu } h^{\mu } (U\partial)
e^{\nu}] \},
\end{eqnarray}
where $({\gamma}^{\prime}_E)_{\mu\nu\rho}$ and
$({\gamma}^{\prime}_M)_{\mu \nu\rho}$ are the gyration
pseudotensors, $(U\partial)=U_{\rho}\partial^{\rho}$.

If pseudotensors
$({\gamma}^{\prime}_E)_{\mu\nu}$ and $({\gamma}^{\prime}_M)_{\mu\nu}$
are determined via $g_ {\mu\nu}$ as
$({\gamma}^{\prime}_{E,M})_{\mu\nu}=({\gamma}^{\prime}_{E,M})
g_ {\mu\nu}$, then  it reduce to violation of the
spatial parity conservation low ({\it i.e.} spin of a composite
particle is not taken into account).

From the expression
(\ref{h24}) it  follows that the usual polarizabilities ($\alpha$,
$\beta$) in a rest medium  and the gyration give non-zero
contributions to the Lagrangian, starting with the second
and the third order in povers of frequency of an external
  electromagnetic field respectively.

Due to the correspondence principles \cite {12} and
the expression (\ref{h24}),
the field theory effective Lagrangian of electromagnetic field
interaction with spinless hadrons will not satisfy
to the parity conservation law when the
components of pseudotensors
${({\gamma}^{\prime}_E)}_{\mu\nu}$ and
${({\gamma}^{\prime}_M)}_{\mu\nu}$
 are not equal to zero.
However, in the case of the spin particles it is easy to determine
dependence of
$\gamma$-structures from $
( \stackrel{-}{\psi
}(\stackrel{\leftrightarrow }{\partial _\alpha }
\stackrel{\leftrightarrow }{\partial _\beta })\gamma^{\mu}\gamma^{5}\psi )$ and
$(F^{\alpha
\nu}\stackrel{\leftrightarrow } { \partial ^\nu }\tilde
{F}_{\sigma \mu }) $.

Hence, the effective field Lagrangian discribing
the electromagnetic field and the spin-1/2 hadron interaction
is defined as follows:
\begin{equation}
\label{h26}
L_{eff}= L_{eff}^{pol}+L_{eff}^{Sp},
\end{equation}
where $L^{pol}_{eff}$  is determined by the expression
(\ref{h21}) and  $L_{eff}^{Sp}$ looks like
\widetext
\begin{eqnarray}
\label{h27}
L_{eff}^{Sp}&=&\frac \pi {4m^2}
( \stackrel{-}{\psi
}(\stackrel{\leftrightarrow }{\partial _\alpha }
\stackrel{\leftrightarrow }{\partial _\beta }+\stackrel
{\leftrightarrow }{\partial _\beta }\stackrel{\leftrightarrow }
{\partial _\alpha })\gamma^{\mu}\gamma^{5}\psi )
\nonumber \\&&
\times \{-\frac 1{2}~\gamma _{E_1}\cdot~
F^{\alpha
\nu}\stackrel{\leftrightarrow } { \partial ^\beta }\tilde {F}_{\mu
\nu }
\nonumber \\&&
+\frac 1{2}~\gamma _{M_1}\cdot~ \tilde {F}^{\alpha
\nu}\stackrel{\leftrightarrow } { \partial ^\beta }F_{\mu \nu }
\nonumber \\&&
-\gamma _{E_2}\cdot~
(F^{\alpha \nu}\stackrel{\leftarrow }{\partial _\mu
}\tilde{F}^{\beta}_ {~\nu }-
\tilde{F}^{\alpha \nu}\stackrel{\rightarrow }{\partial _\nu
}F_{\mu}^ {~\beta })
\nonumber \\&&
+\gamma _{M_2}\cdot~ (\tilde{F}^{\alpha \nu}\stackrel{\leftarrow }{\partial _\mu
}F^{\beta}_ {~\nu }-
F^{\alpha \nu}\stackrel{\rightarrow }{\partial _\nu
}\tilde{F}_{\mu}^ {~\beta })\},
\end{eqnarray}
\narrowtext
here
$
\gamma ^5=-\left( \frac 0{I} \frac I{0}\right) $.

 The effective interaction Hamiltonian of the
electromagnetic field and the spin-1/2 hadron  follows
from the Lagrangian ~(\ref{h27}) \cite{7} \begin{eqnarray}
H_{eff}^{Sp}&=&-2\pi ( \gamma _{E_1} {\mbox{\boldmath $ \sigma$}
}\cdot {\bf E}\times \stackrel{\bullet }{\bf E}+ \gamma _{M_1}
{\mbox{\boldmath $ \sigma$}}\cdot {\bf H}\times \stackrel{\bullet }{
\bf H} \nonumber \\&& - 2\gamma _{E_2}E_{ij}\sigma _iH_j+
\label{h28} 2\gamma _{M_2}H_{ij}\sigma _iE_j), \end{eqnarray} where
$E_{ij}=
\frac1{2}(\partial_{i}E_{j}+\partial_{j}E_{i})$ and  $H_{ij}=
 \frac1{2}(\partial_{i}H_{j}+\partial_{j}H_{i}).$

The effective  Lagrangian (\ref{h26}) satisfies both the crossing
symmetry and all requirements of the relativistic quantum field
theory. As we can see from (\ref{h21}), at the ${\cal O}(\omega^2)$
the effective Lagrangian
of spin-1/2 hadrons are determined by the electric and magnetic
polarizabilities only ($\alpha _0$ and $\beta _0$).

Besides, as
follows
from expression ~(\ref{h27}), spin polarizabilities for the
spin-1/2 hadrons give the contributions to the effective Lagrangian
which are of order  ${\cal O}(\omega^3)$.
Using the Lagrangian (\ref{h27}) the Compton scattering
amplitude for the spin-1/2 hadrons have been obtained:
\begin{eqnarray}
\label{h29}
\frac 1{4\pi m}T_{fi}^{NB, spin}=-\chi ^{+}
[-i~ &\gamma _{E_1}&~ \mbox{\boldmath $
\sigma$} \cdot \bf {e}^{\prime * }\times \bf {e}(\omega _1^2\omega
_2+\omega _1\omega _2^2)
\nonumber \\
-i~ &\gamma _{M_1}&~
\mbox{\boldmath $ \sigma$}\cdot \bf {s}^{\prime * }\times \bf
{s}(\omega _1^2\omega _2+\omega _1\omega _2^2)
\nonumber \\
-i~&\gamma _{E_2}&~\{ [\mbox{\boldmath $ \sigma$}\cdot \bf {n}
\bf {e}\cdot \bf {s}^{\prime *}+\mbox{\boldmath $ \sigma$}\cdot\bf
{e} \bf {n}\cdot \bf {s}^{\prime*}]\omega _1^2\omega _2
\nonumber\\
&&-[\mbox{\boldmath $ \sigma$}\cdot \bf
 {n}^{\prime} \bf {e}^{\prime *}\cdot \bf {s}+ \mbox{\boldmath $
\sigma$}\cdot\bf {e}^{\prime *} \bf {n}^{\prime}\cdot \bf
{s}]~\omega _1\omega _2^2\}
\nonumber \\
+i~&\gamma _{M_2}&~\{ [\mbox{\boldmath $ \sigma$}\cdot \bf {n}
\bf {e}^{\prime *}\cdot \bf {s}+ \bf {n}\cdot\bf {e}^{\prime *}
\mbox{\boldmath $ \sigma$}\cdot \bf {s}]\omega _1^2\omega _2
\nonumber \\
&&-[\mbox{\boldmath $ \sigma$}\cdot \bf {n}^{\prime}
\bf {e}\cdot \bf {s}^{\prime *}+\bf {n}^{\prime} \cdot \bf
 {e} \mbox{\boldmath $ \sigma$}\cdot \bf {s}^{\prime *}]\omega_1
\omega _2^2\}]\chi.
\end{eqnarray}
The amplitude (\ref{h29}) for Compton scattering
 contains the structures typical for the spin
polarizabilities and satisfies both the gauge invariance and the
crossing symmetry.

To summarize,
on the basis of the correspondence principle between the relativistic
moving medium electrodynamics and relativistic quantum field theory
the covariant Lagrangian of the electromagnetic field interaction
with the polarized spin particles have been obtained. This
Lagrangian satisfies the main relativistic quantum field theory
requirements (is cross-invariant, P and
T- and gauge invariant). The obtained Lagrangian can be
used for description of spin polarizabilites in two-photon processes
(Compton scattering, scattering of electrons  on nucleones, and
others).

\end{document}